\documentclass[aps,pra,twocolumn,groupedaddress,longbibliography]{revtex4-1}
\usepackage{graphicx}
\usepackage{caption}
\DeclareCaptionLabelSeparator{dot}{. }
\makeatletter
\def\justified{
	\let\\\@normalcr
	\@rightskip\z@skip \rightskip\@rightskip
	\leftskip\z@skip
	\parindent 0em\relax
	\setlength{\parfillskip}{0pt plus 1fil}}
\DeclareCaptionJustification{justified}{\justified}
\usepackage{subcaption}
\usepackage{amsmath}
\usepackage{amssymb}
\usepackage{mathrsfs} 
\usepackage{braket}
\usepackage{units}
\usepackage{ragged2e}
\usepackage[colorlinks,urlcolor=blue ,citecolor=blue ,linkcolor=blue ]{hyperref}
\usepackage{xcolor}
\usepackage{nicefrac} 
\usepackage{lipsum}
\usepackage{nameref}
\usepackage{hyperref}
\usepackage{textcomp} 
\usepackage{urwchancal}
\usepackage{xcolor}
\usepackage{float}
\definecolor{darkgreen}{rgb}{0,0.5,0}
\usepackage{multirow}
\usepackage{tabularx}
\usepackage[normalem]{ulem}

\newcommand{\bs}{\boldsymbol}

\newcommand{\vB}{\ensuremath{\bs{B}}}
\newcommand{\vk}{\ensuremath{\bs{k}}}

\newcommand{\br}{\ensuremath{\bs{r}}}

\newcommand{\ac}{\ensuremath{a_{\rm c}}}
\newcommand{\accrot}{\ensuremath{a_{\rm rot}^*}}
\newcommand{\acc}{\ensuremath{a^*}}
\newcommand{\as}{\ensuremath{a_{s}}}
\newcommand{\add}{\ensuremath{a_{\rm dd}}}
\newcommand{\edd}{\ensuremath{\epsilon_{\rm dd}}}
\newcommand{\gs}{\ensuremath{g_{s}}}
\newcommand{\gdd}{\ensuremath{g_{\rm dd}}}

\newcommand{\Erot}{\epsilon_{\rm rot}}
\newcommand{\lrot}{\lambda_{\rm{rot}}}
\newcommand{\krot}{k_{\rm{rot}}}

\newcommand{\Erb}{\ensuremath{^{166}}{\rm Er}}
\newcommand{\Dyb}{\ensuremath{^{164}}{\rm Dy}}
\newcommand{\Dybb}{\ensuremath{^{162}}{\rm Dy}}

\newcommand{\Vint}{V_{\rm{int}}}
\newcommand{\Vdd}{V_{\rm {dd}}}
\newcommand{\Vc}{V_{\rm {c}}}
\newcommand{\tVint}{\tilde{V}_{\rm{int}}}
\newcommand{\tVdd}{\tilde{V}_{\rm{dd}}}

\newcolumntype{Y}{>{\centering\arraybackslash}X}

\begin{document}
	
    \title{Quantum-stabilized states in magnetic dipolar quantum gases}
    
	\author{Lauriane Chomaz$^{1}$.}
	
	\affiliation{%
		$^{1}$ Physikalisches Institut, Universität Heidelberg, Im Neuenheimer Feld 226, 69120, Heidelberg, Germany. chomaz@uni-heidelberg.de}	

	\date{\today}
	
\begin{abstract}
 A decade ago, a universal stabilization mechanism driven by quantum fluctuations was discovered in ultracold Bose gases of highly magnetic atoms. This mechanism prevents these systems from collapsing and instead allows exotic states of matter to arise, including ultradilute quantum droplets, crystallized quantum states, and specifically supersolids. We review the experimental and theoretical progress in understanding these quantum-stabilized states, their emergence, and intriguing properties.

 \textbf{keywords:} quantum gases, dipolar interactions, quantum-stabilization, superfluids, droplets, supersolids
\end{abstract}

\maketitle

\section{Introduction}

Ultracold gases offer a prime platform for exploring quantum phenomena with exceptional control. Their high purity and tunability enable the engineering of a variety of Hamiltonians by manipulating particle statistics, trapping potential, internal atomic structure, and interparticle interactions~\cite{Bloch2008mbp,Pitaevskii2016bec}. While most atomic species interact via isotropic contact interactions of tunable strength, introducing long-range and anisotropic interactions has been highly desirable to enrich the physics at reach~\cite{Baranov2008,Baranov2012cmt}. 

A naturally occurring long-range interaction between neutral particles is the dipole-dipole interaction. This interaction can be realized in ultracold gases using atoms with large magnetic moments in their ground state~\cite{Chomaz2023dipolar},  polar molecules with induced electric dipoles~\cite{Bohn2017cmp}, or atoms excited to Rydberg states~\cite{Browaeys2020mbp}.  Each of these systems operates in distinct parameter regimes and faces different experimental challenges. 

Here we focus on highly magnetic atoms, for which large assemblies of tens of thousands of atoms are routinely brought to quantum degeneracy. These assemblies exhibit long lifetimes and readily competing interactions of different natures. 
Chromium atoms formed the first magnetic quantum gas in 2005~\cite{Griesmaier2005bec}. 
Significant advancements came with the realization in the early 2010s of quantum gases of even more magnetic open-shell lanthanide atoms~\cite{Lu2011sdb,Aikawa2012a}. Despite initial concerns about their cooling and trapping, these species have proven remarkably accessible and highly tunable, leading to the rapid development of many experimental platforms~\cite{Chomaz2023dipolar}.

A turning point in the field came in 2015 with the discovery of an unforeseen stabilization mechanism -- induced by the very effects of quantum fluctuations -- that prevents magnetic quantum Bose gases from collapsing in regimes where standard theory predicts they do~\cite{Kadau2016otr}. Based on this mechanism and its understanding, novel states of matter have been discovered in the following years, including ultradilute liquid droplets, crystallized quantum states, and most notably  supersolids~\cite{Chomaz2016qfd,schmitt2016sbd,Wenzel2017ssi,Tanzi2019ooa,Chomaz2019lla,Bottcher2019tsp}. This article aims to review our current understanding of the wealth of quantum-stabilized states in dipolar Bose gases, focusing on the case of polarized lanthanide atoms. We discuss key experimental achievements, theoretical descriptions and predictions, as well as promising directions for future research.


\section{Interactions in highly magnetic quantum Bose gases}
\label{sec:experiment}

\begin{figure*}[ht!]
\centering
\includegraphics[width=1.\textwidth]{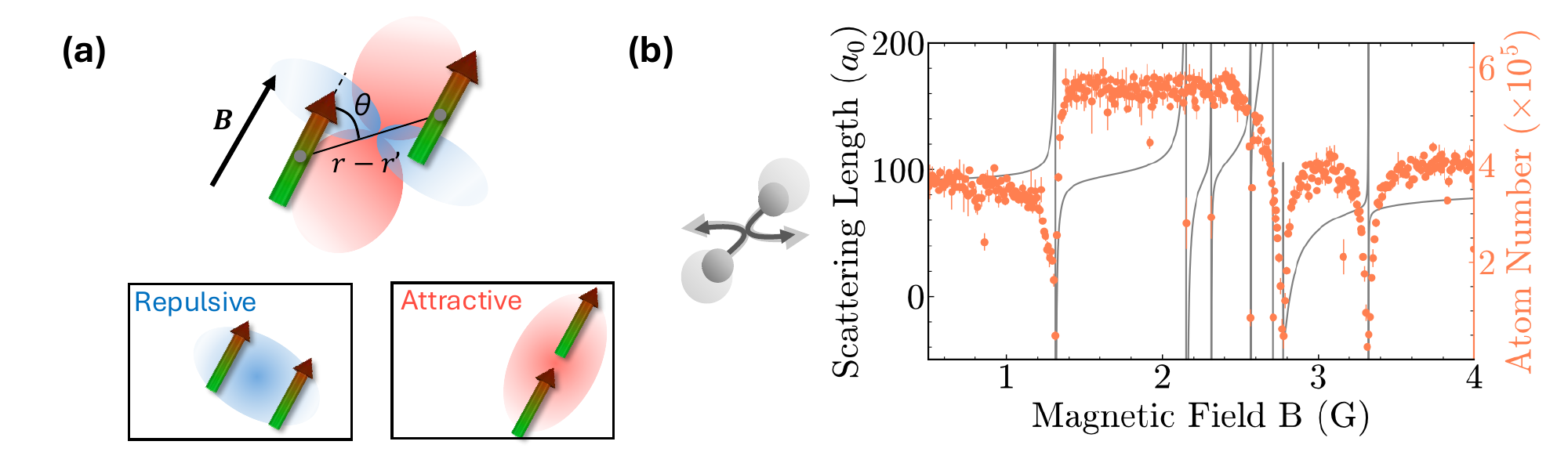}

\caption{{\textbf{Interactions between magnetic atoms.} 
(a) Sketch of the DDI between two magnetic atoms polarized by $\vB$. The interaction is long-range and anisotropic (red and blue halos). It is attractive (red) when the atoms are head to tail (lower right panel) and repulsive (blue) when they are side by side (lower left panel). (b) Left: Sketch of the contact interaction and its tunable strength. Right: Feshbach spectrum measured via atom loss spectroscopy on thermal samples of \Dyb~atoms in the apparatus of Ref.~\cite{Jin2023tdm} for $B$ varying from 0 to 4G, together with the expected variation of $\as$. At each resonance $\as$ diverges. The background scattering length is set to $85.5\,a_0$, see e.g.~\cite{Chomaz2023dipolar}.
}} 
\label{fig:interaction_sketch}
\end{figure*}


\subsection{Dipole-dipole interactions}

Atoms with a large magnetic moment $\mu$ in their ground state interact via sizeable dipole-dipole interactions (DDI). 
In the presence of an external magnetic field $\vB$, the dipoles get spontaneously polarized and the magnetic DDI is
\begin{equation}
\label{eq:vdd}
\Vdd(\br)=\frac{3\gdd}{4 \pi r^3} \left[1-3\cos^2\theta \right],
\end{equation}
where $\theta$ is the angle between the interparticle position $\br$ and the polarizing field $\vB$ (i.e.~the dipole direction)~\cite{Baranov2008,Chomaz2023dipolar}. The DDI is anisotropic and changes sign, being attractive (resp.~repulsive) when the dipoles are head-to-tail (resp.~side-by-side) [$\theta=0$ (resp.~$\pi/2$)], see Fig.~\ref{fig:interaction_sketch}(a).   The strength of the dipolar interaction is parametrized by the coupling constant $\gdd=4 \pi \hbar^2 \add/m$ scaling with the dipolar length $\add = m\mu_{\rm B}\mu^2/12\pi\hbar^2$. Here $m$ is the atomic mass, $\hbar$ the reduced Planck constant, and $\mu_{\rm B}$ the Bohr magneton. In this review, we focus on open-shell lanthanide atoms, specifically Er and Dy, for which  $\add \approx 66\,a_0$ and $130\,a_0$, respectively~\cite{Chomaz2023dipolar}. 

The $1/r^3$ scaling of \eqref{eq:vdd} decays sufficiently slowly with $r$ so that the DDI remains long-range and anisotropic with all partial waves contributing to the scattering in the ultracold dilute regime relevant to quantum gases~\cite{Pitaevskii2016bec,Chomaz2023dipolar,Baranov2008,Baranov2012cmt}. 
This contrasts with the van der Waals interaction potential, scaling as $1/r^6$, for which only the $s$-partial wave remains contributing in the ultracold dilute regime~\cite{Pitaevskii2016bec}. 

\subsection{Tunable contact interactions}

Van der Waals forces dominate the interactions between ultracold non-magnetic atoms. They also exist between magnetic atoms, giving rise to a second interaction term in addition to $\Vdd$. In the ultracold dilute regime, this term is well captured by a contact pseudopotential
\begin{equation}
\label{eq:vc}
\Vc(\br)=\gs \delta(\br),
\end{equation}
where the coupling constant $\gs={4 \pi \hbar^2 \as}/{m}$ is set by the $s$-wave scattering length $\as$~\cite{Pitaevskii2016bec}. 
In contrast to $\Vdd$, $\Vc$ is isotropic and short-range. The total pseudopotential is given by $\Vint(\br) = \Vdd (\br)+\Vc(\br)$ with $\as$ including the dipolar $s$-wave contribution~\cite{Yi2000tac,Yi2001a,Bortolotti2006sli,Baranov2008}. The ratio $\edd=\add/\as$ characterizes the contact-DDI competition.

Interestingly, $\as$ can be widely tuned experimentally by altering the magnitude of $\vB$ and exploiting Feshbach resonances~\cite{Pitaevskii2016bec,Bloch2008mbp}. Magnetic lanthanide atoms exhibit remarkably dense Feshbach spectra with multiple resonances addressable with an offset field of just a few Gauss, see Fig.~\ref{fig:interaction_sketch}(b)~\cite{Baumann2014ool,Frisch2014qci,Maier2015eoc,Chomaz2023dipolar}. 
This enables the use of easy-to-generate weak magnetic fields, which in turn provides remarkable tunability with both $\as$ and the dipole orientation being readily adjusted through several pairs of small coils controlling $\vB$~\cite{Chomaz2023dipolar}.

\section{Basic many-body aspects of dipolar quantum Bose gases}\label{sec:theory}

\subsection{Mean-field description of dipolar gases}\label{subsec:MF}


At ultralow temperatures, due to statistical effects, bosons form a macroscopic wavefunction extending over the whole system - a Bose-Einstein condensate (BEC)~\cite{Pitaevskii2016bec}. Deep in this quantum degenerate regime, weakly interacting Bose gases can be described by a classical complex field, $\psi(\mathbf{r})$. It embodies the macroscopic wavefunction but neglects its fluctuations, and interactions are accounted for only at the mean-field (MF) level. Within this framework, 
$\psi(\mathbf{r})$ satisfies the Gross-Pitaevskii equation (GPE), which for dipolar gases writes
      \begin{equation}
        \label{eq:GPE}
{\mathrm i}\hbar\frac{\partial\psi}{\partial t}=\left[\frac{-\hbar^2\Delta}{2m}+V(\br)+\gs n(\br)+\Phi_{\rm dd}(\br)  \right]\psi.
\end{equation}
Here $n(\br)=|\psi(\br)|^2$ is the spatial density, $\psi(\br)$ is normalized to the total particle number $N=\int |\psi(\br)|^2d^3 r$~\cite{Pitaevskii2016bec}. $V(\br)$ represents the trapping potential. The last two terms arise from the interaction potential $\Vint(\br)$. The first term is the local contact term, and the second term 
\begin{equation}
        \label{eq:phidd}
\Phi_{\rm dd}(\br)=\int \Vdd(\br-\br')n(\br')d^3 r'
\end{equation}
 is the non-local convolution of the DDI potential \eqref{eq:vdd} with the density~\cite{Baranov2008}. 
 The GPE describes the degenerate-gas dynamics, by solving Eq.\,\eqref{eq:GPE}
 , as well as its ground state, by replacing the left-hand side of \eqref{eq:GPE}, ${\mathrm i}\hbar\frac{\partial\psi}{\partial t}$, by $\mu_0 \psi$, where $\mu_0$ is the gas chemical potential. 

\subsection{Mean-field dipolar effects --- superfluid excitation spectrum}\label{subsec:Vk}

\begin{figure*}[ht!]
\centering
\includegraphics[width=1.\textwidth]{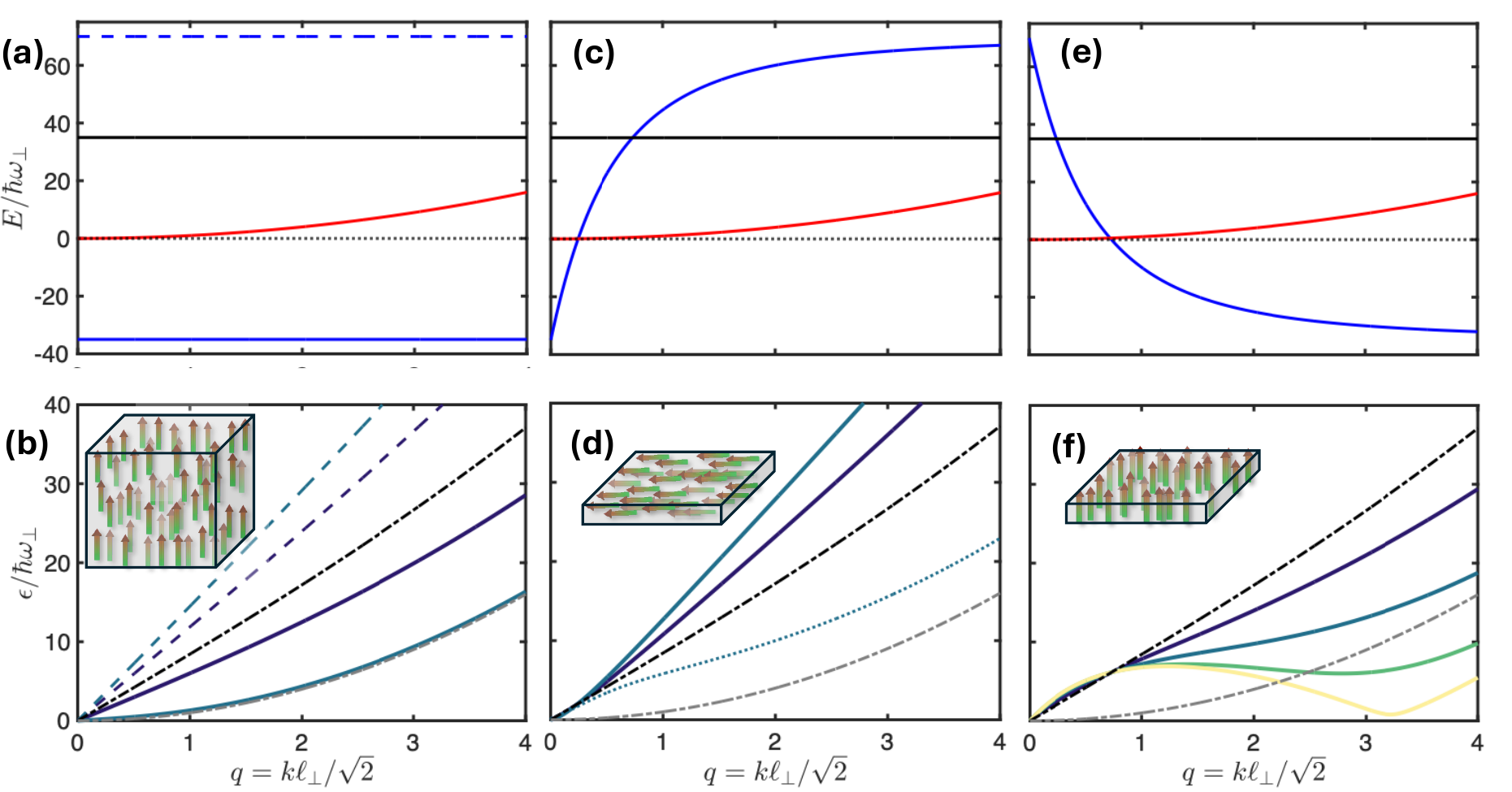}

\caption{{\textbf{Interactions and dispersion relation of a dipolar gas} (a,b): for a 3D unconfined gas, (c-f): in an infinite pancake trap with dipoles (c,d) along, (e,f) transverse to the excitation direction. The geometries are illustrated as insets of (b,d,f). The energies are rescaled by $\hbar\omega_{\perp}=h \times 150\,$Hz and the momenta by $\sqrt{2}/\ell_\perp$. The parameters are chosen so that $\gs n=35\hbar\omega_\perp$ and matching \Dyb~values, namely $\as=130.8a_0$, $m=164\,$atomic mass units, $n_{3D}=9.8\times10^{20}m^{-3}$. The pancake transverse confinement frequency is set to $\omega_{\perp}$. We use a quasi-2D approximation with the gas transverse distribution set to the harmonic oscillator ground state, $n_{2D}=\sqrt{2\pi}n_{3D}\ell_\perp$, and neglect magnetostriction, which quantitatively modifies $\tVdd(\vk)$ and $\epsilon(\vk)$. Qualitative behaviors are illustrated.  (a,c,e): Characteristic energies:  contact interaction energy $\gs n$ (solid black), DDI energy $\tVdd(\vk) n$ for $\edd=1$ (blue), kinetic energy $\hbar^2 k^2/2m$ (red). (b,d,f): Dispersion relation \eqref{eq:Bdg} for $\edd=0.5$ (dark blue), $\edd=1$ (light blue), $\edd=1.25$ (green), $\edd=1.31$ (yellow) together with the non-dipolar case $\edd=0$ (black dashed-dotted) and the non-interacting one $\gs=\gdd=0$ (gray dashed-dotted). The higher $\edd$ values are shown only in (f) and unstable elsewhere. (a,b) show two excitation directions: $\theta_k=0$ (dashed lines) and $\theta_k=\pi/2$ (solid lines). In (d), the antiroton behavior expected beyond the longitudinal approximation is sketched (dotted blue line, inspired from Ref.~\cite{Pal2020}).
}} 
\label{fig:dispersion}
\end{figure*} 

The non-local and anisotropic nature of the DDI gives rise to remarkable features in degenerate dipolar gases. These features can be scrutinized by considering the interaction potential in momentum space, $\tVint(\vk)$. In particular, $\tVint(\vk)$ provides insights into the elementary excitation spectrum, which is key to understanding the gas' dynamical and thermodynamic behaviors~\cite{Pitaevskii2016bec,Bloch2008mbp}.

\subsubsection{Fully unconfined gases}\label{subsec:unconfexc}

For a fully unconfined gas, the elementary excitations are waves of momentum $\vk$, i.e.~wavelength $2\pi/|\vk|$. Within Bogoliubov theory -- applicable to weakly interacting gases -- the energies $\epsilon(\vk)$ of these excitations assume the form:
\begin{eqnarray}
\epsilon(\vk)&=& \sqrt{\frac{\hbar^2 k^2}{2m}\left(\frac{\hbar^2 k^2}{2m}+2\tVint(\vk)n\right)},
\label{eq:Bdg}
\end{eqnarray}
with $n$ the ground-state density-- here assumed to be uniform~\cite{Chomaz2023dipolar,Baranov2008,Baranov2012cmt,Pitaevskii2016bec}.

For a contact-interacting gas [$\Vint(\br)=\Vc(\br)$], the momentum-space interaction potential is given by the Fourier transform of Eq.~\eqref{eq:vc}, $\tVint(\vk) = \gs$, and is constant across all momenta, see Fig.\,\ref{fig:dispersion}(a). For such a potential, Eq.~\eqref{eq:Bdg} evolves from a linear phononic dispersion at small $k$, $\epsilon(k)= \hbar ck$ with $c=\sqrt{{\gs n}/{m}}$ the speed of sound, to a quadratic free-quasi-particle dispersion $\epsilon(k) \sim \hbar^2 k^2/2m$ at large $k$, see Fig.\,\ref{fig:dispersion}(b)~\cite{Pitaevskii2016bec}.

For a uniform and unconfined three-dimensional (3D) dipolar gas, the momentum-space DDI is also simply given by the Fourier transform of Eq.\,\eqref{eq:vdd}, 
\begin{eqnarray}
\tVdd(\vk) = \gdd \left[3 \cos(\theta_k)^2 -1 \right], 
\label{eq:vddtf}
\end{eqnarray}
where $\theta_k$ is the angle between $\vk$ and the dipole axis~\cite{Chomaz2023dipolar,Baranov2008}. The full momentum-dependent interaction is then $\tVint(\vk) = \gs+\tVdd(\vk)$. 

Remarkably, $\tVint(\vk)$ does not depend on the magnitude of $\vk$ but only on its orientation, see Fig.\,\ref{fig:dispersion}(a). Thus, the DDI does not introduce new types of elementary excitations here. Instead, it induces anisotropic features in the excitation spectrum and speeds of sound, see Fig.\,\ref{fig:dispersion}(b). Specifically, the modes along the dipole direction ($\theta_k=0$) harden (increased $\epsilon(\vk)$ and $c$), while transversely ($\theta_k=\pi/2$) they soften (reduced $\epsilon(\vk)$ and $c$). This additionally shifts the threshold scattering length $\acc$ at which phonons become unstable (i.e., where the speed of sound vanishes), from $\acc=0$ in the purely contact-interacting case to $\acc=\add$ (i.e.~$\edd=1$), see Fig.\,\ref{fig:dispersion}(b) and Eq.~\eqref{eq:vddtf} at $\theta_k=\pi/2$~\cite{Baranov2008,Baranov2012cmt}.  

\subsubsection{Role of Confinement}\label{subsec:confexc}

A confining potential $V(\br)$, especially if anisotropic, modifies this picture and can lead to new features in the dispersion relation $\epsilon(\vk)$. Intuition can be gained by considering a gas confined only in one or two spatial directions, with harmonic trapping frequencies $\omega_\perp$, but unconfined in the excitation direction $\vk$ (i.e.~infinite tube or pancake).

Assuming that excitations in the unconfined direction are purely longitudinal, i.e.~retain the ground-state's shape in the transverse confined directions, 
Eq. \eqref{eq:Bdg} remains valid. 
Yet, $n$ becomes the linear or areal density and $\tVint(\vk)$ is no longer the Fourier transform of $\Vint(\br)$. Instead, its dipolar contribution $\tVdd(\vk)$ relates to $\Phi_{\rm dd}(\br)$ [Eq.~\eqref{eq:phidd}] and contains information on $n(\br)$ spatial variations induced by the confinement~\cite{Santos2003rms,Fischer2006a,Ronen2006bmo, Sinha2007cdg, Baillie2015a, Chomaz2018oot, Blakie2020vtf, Pal2020,Baranov2008}. Thereby, $\tVdd(\vk)$ acquires a dependence on $k=|\vk|$, potentially giving rise to elementary excitations of neither phononic nor free-particle nature, see Fig.\,\ref{fig:dispersion}(c-f). 

The characteristic momentum scale for $\tVdd$ variations with $k$, and thereby for the novel excitation features, is set by the inverse confinement length $\ell_{\perp}^{-1}$ with $\ell_{\perp}=\sqrt{\hbar/m\omega_\perp}$. These variations are qualitatively dictated by the relative orientation of the dipoles, confined direction(s), and excitation direction $\vk$, as reviewed below.

\paragraph{Confinement perpendicular to the dipoles ---}\label{par:excitation_phonons}
 If the dipoles are oriented along the excitation direction, perpendicular to the confinement direction(s),  $\tVdd$ increases with $k$, from negative to positive values, see Fig.\,\ref{fig:dispersion}(c)~\cite{Sinha2007cdg,Pal2020}. This yields a hardening of the large-$k$ modes, see Fig.\,\ref{fig:dispersion}(d). The resulting high excitation energies cause a breakdown of the longitudinal approximation of Eq.\,\eqref{eq:Bdg} once $\epsilon(\vk) \sim \hbar\omega_\perp$. Going beyond this approximation and accounting for the transverse reshaping of the cloud under excitation gives rise to a novel \emph{antirotonic} feature in the dispersion relation through an avoided level crossing, see Fig.\,\ref{fig:dispersion}(d)~\cite{Pal2020,Houwman2024mot}. 

Conversely, the DDI causes a softening of the phonon modes, which remain stable (i.e.~real-valued $\epsilon(\vk)$) as long as $\edd \leq 1$ thanks to the additive contact term $\gs$, see Fig.\,\ref{fig:dispersion}(d). As in the unconfined case, the phonon-instability threshold shifts to $\acc=\add$ (i.e.~$\edd=1$). 

\paragraph{Confinement along the dipoles ---}\label{par:excitation_rotons}
If instead the dipoles are oriented along a confinement direction, perpendicular to the excitation direction, $\tVdd$ decreases with $k$, starting positive at $k=0$ and becoming negative (attractive) at large $k$, see Fig.\,\ref{fig:dispersion}(e)~\cite{Santos2003rms,Chomaz2018oot,Fischer2006a, Blakie2020vtf, Pal2020}. This hardens long-wavelength, and softens short-wavelength excitations, see Fig.\,\ref{fig:dispersion}(f). Here, the longitudinal approximation of Eq.\,\eqref{eq:Bdg} remains reliable even for shallow traps and for $\epsilon(\vk)>\hbar\omega_\perp$~\cite{Pal2020}.  Due to phonon-modes hardening, the DDI may exceed the contact interaction in strength (i.e.~$\add>\as$, $\edd>1$) without the spectrum becoming unstable (i.e.~real-valued $\epsilon(\vk)$), see Fig.\,\ref{fig:dispersion}(f). For dominant DDI, the competition in \eqref{eq:Bdg} between kinetic term $\hbar^2 k^2/2m$ and interactions with attractive character at large $k$ ultimately yields new excitation features. Namely, the dispersion develops a local maximum (maxon) followed by a local minimum (roton), see Fig.\,\ref{fig:dispersion}(f) for $\edd=1.25$ and $1.31$~\cite{ODell2003rig,Santos2003rms,Chomaz2018oot,Fischer2006a, Blakie2020vtf, Pal2020}.

A similar dispersion shape was first postulated by Landau (and later observed) to explain the distinctive dynamical properties of superfluid liquid helium~\cite{Landau1947ott}. The possibility of a maxon-roton dispersion is one of the most striking features of dipolar superfluids. It signals that density modulations at a specific wavelength -- that of the roton $\lrot=2\pi/\krot$ -- are energetically favored~\cite{Feynman1954, Nozieres2004itr}. The dipolar roton minimum arises at the MF level, and $\lrot \sim \ell_{\perp}$ with additional dependencies on density and interaction strengths, see Fig.\,\ref{fig:dispersion}(f). 

The dipolar roton energy gap $\Erot$ can be finely adjusted 
by tuning $\as$, as $\gs$ acts as an offset in $\tVint(k)$, see \ref{subsec:unconfexc}. $\Erot$ decreases with $\as$ down to $\as=\accrot$, for which the roton is the first mode of the spectrum to fully soften, i.e.~$\Erot=0$, see Fig.\,\ref{fig:dispersion}(f). The existence of a roton minimum and its controlled softening was tested in two experiments using \Erb~atoms, probing the dispersion via Bragg scattering, and the growth of a coherent population in the roton mode after an interaction quench past its full softening, respectively~\cite{Petter2019ptr,Chomaz2018oot}. Additional confirmations came from density fluctuations analysis in \Dybb~quantum gases~\cite{Hertkorn2021dfa,Schmidt2021rei}.

\paragraph{Fully trapped case ---}\label{par:fulltrapexc}

Going beyond the approximation of \eqref{eq:Bdg}, the dispersion relation and the underlying elementary excitations can be derived within Bogoliubov theory, linearizing the GPE~\eqref{eq:GPE} around the ground-state wavefunction~\cite{Pitaevskii2016bec,Baranov2008}. This in particular holds for a fully trapped system. We note $\omega_{\parallel}$ and $\omega_\perp$ the longitudinal and tightest transverse trap frequencies, respectively. While a 3D trap alters the excitation spectrum properties -- e.g. making it discrete, or causing a momentum broadening in non-uniform traps -- many features derived in \ref{par:excitation_phonons},\ref{par:excitation_rotons} ($\acc$ shift, DDI-induced softening and hardening) remain relevant. For large trap aspect ratio $\omega_\perp/\omega_{\parallel}$, the antiroton and roton-maxon features themselves are retained, while for smaller aspect ratios, the interplay of trap geometry and DDI persists and yields excitations of peculiar character, see also \ref{subsec:trappedPD}~\cite{Ronen2006bmo,Ronen2007raa,Blakie2012rsi,Martin2012a,JonaLasinio2013rci,Bisset2013rei,Chomaz2023dipolar,Chomaz2018oot,Houwman2024mot,Hertkorn2021dfa,Schmidt2021rei,Petter2019ptr}. The formalism of \eqref{eq:Bdg} thus stays insightful for experimentally relevant configurations.

\subsection{Dipolar quantum gases beyond the mean-field instability}\label{subsec:BMF}

The ability to fully soften the roton mode -- bringing its energy $\Erot$ to equate that of the superfluid ground state -- raises the question of what the ground state is at and beyond this point. This question was first posed in the context of superfluid and solid helium, where it sparked vivid debates for decades~\cite{Penrose1956bec,Gross1957uto, Boninsegni2012csw,Yukalov2020}. With the advent of ultracold gases, especially dipolar ones, this question was revived
~\cite{Giovanazzi2002dmo, ODell2003rig,Santos2003rms, Komineas2007vli, Buchler2007a, Boninsegni2012spo, Lu2015a, Yukalov2020, Leonard2017sfi, Li2017asp}. 

However, after early debates~\cite{Yi2001a,Santos2003rms,Fischer2006a,Ronen2007raa,Baranov2008}, before 2015, it was commonly believed that when the roton mode of a dipolar gas dynamically reaches full softening, e.g. upon tuning $\as$ below $\accrot$, the system undergoes an abrupt dynamics resulting in the rapid loss of the BEC -- the collapse~\cite{Kagan1998ceb, Sackett1998a, Komineas2007vli, Koch2008a,Lahaye2009dwc,Bohn2009hda,Parker2009sfd}. This collapse occurs due to the dominance of attractive interactions at some finite momenta, triggering a sharp local density increase, which, in turn, leads to strong losses via three-body processes. It resembles the collapse occurring in contact-interacting gases or dipolar gases confined in traps elongated along the dipole direction. Yet, in the latter cases, the collapse is driven by the softening of the $k\rightarrow 0$ phonon modes and occurs, if adiabatic, through a global density increase (see also discussion in \ref{subsec:modulation}).  Within the MF theory described in \ref{subsec:MF}, no ground states of finite density exist for $\as<\accrot$, providing the static counterpart to the dynamical instability described above.

At the end of 2015, a groundbreaking discovery modified this prevailing understanding.  An experiment revealed the surprising survival of the degenerate gas of \Dyb~on long time scales after quenching to the MF instability regime~\cite{Kadau2016otr}. Furthermore, tiny density structures were formed in the gas. At first, the origin of the stabilization was not understood~\cite{Kadau2016otr,Blakie2016a,Bisset2015coa,Xi2016a}. Soon after, however, a parallel was drawn to a recent prediction that contact-interacting mixtures could be stabilized through beyond-MF effects, i.e.~related to the quantum fluctuations of the field $\psi$~\cite{Petrov2015qms,FerrierBarbut2016ooq}.

 In a perturbative picture, beyond-MF effects give rise to a correction to the energy and chemical potential of the degenerate gas beyond the MF terms comprised in the brackets of Eq.\,\eqref{eq:GPE}. The leading-order correction is given by the Lee Huang Yang (LHY) term, which for dipolar gases reads 
 \begin{equation}
        \label{eq:LHY}
 \mu_{\rm LHY} (n,\edd)=\frac{32}{3\sqrt{\pi}}gn\sqrt{n\as^3}F(\edd)\equiv \gamma_{\rm QF}n^{3/2}.
\end{equation}
 This term acts as an effective repulsive interaction term with steeper scaling with $n$ than MF (see two last terms in Eq.\,\eqref{eq:GPE}). The function $F(\edd)$ is given by 
$F(\edd)=\frac{1}{2}\int d\theta_k \sin{\theta_k} (1+\edd(3\cos^2\theta_k-1))^{5/2}$~\cite{Schutzhold2006mfe,Lima2011qfi,Lima2012bmf}. The above formula is derived for a uniform unconfined gas and for $\edd<1$ but is commonly directly used for $\edd\geq 1$ and for trapped systems through local density approximation. This is justified as $\mu_{\rm LHY}$ is dominated by the contribution of large-$k$ modes, corresponding to small spatial structures compared to the trap size. These modes are also hard modes and insensitive to the phonon instability occurring at $\edd\geq 1$ in unconfined systems (see \ref{subsec:unconfexc}). This instability instead yields a small imaginary part in $\mu_{\rm LHY}$ due to the soft modes' contribution, which is discarded~\cite{Wachtler2016qfi,Bisset2016gsp,Saito2016pim}.

While this correction is usually small in weakly interacting gases with $\sqrt{n\as^3}\ll1$, it becomes important in the special case where MF interactions nearly cancel while $\as$ remains finite. This is exactly what can happen in magnetic gases, as the DDI competes with the short-range interaction and can cancel its contribution at MF level with $\as\sim \add \neq 0$
~\cite{Wachtler2016qfi,Bisset2016gsp,Saito2016pim,FerrierBarbut2016ooq,Chomaz2016qfd}. In the presence of weak MF attraction, $\mu_{\rm LHY}$ stabilizes states with higher but finite density compared to the MF repulsive case.  This applies both in the dynamics, preventing collapse, and at equilibrium, yielding quantum-stabilized ground states. The density scale is set by \eqref{eq:LHY} and $\as \sim \add$.
Such a stabilization mechanism is universal and applies to all gases, if sufficiently dipolar, regardless of the trap configuration.  Stabilization was not observed in previous studies using chromium~\cite{Lahaye2009dwc,Koch2008a} due to its smaller dipolar length, $\add=15a_0$, which makes three-body losses dominate over the LHY term \eqref{eq:LHY}~\cite{Chomaz2023dipolar}.  

As the gas remains weakly interacting, an effective MF treatment of the beyond-MF stabilization is possible. This is done by adding the LHY term \eqref{eq:LHY} to the bracket in the right-hand side of the GPE \eqref{eq:GPE}~\cite{Wachtler2016qfi,Bisset2016gsp,Saito2016pim,Chomaz2016qfd}. This modified equation is known as the extended GPE (eGPE). 

The discovery of the quantum-fluctuation stabilization mechanism was confirmed by further experiments~\cite{FerrierBarbut2016ooq, Chomaz2016qfd, schmitt2016sbd}. While the states created in the seminal work of Ref.~\cite{Kadau2016otr} were highly excited due to the specifics of the protocol, soon after, ground states of different nature were formed. Hereafter, we review the wealth of quantum-stabilized states observed and predicted in magnetic dipolar quantum gases.

\section{Quantum-stabilized ground states}\label{sec:PD} 


\subsection{Ground state Modulation}\label{subsec:modulation}

The characteristics of ground and excited states stabilized beyond the MF instability depend on the mode that softens and drives the instability. 
While phonon modes always drive the instability in contact-interacting gases, this differs in dipolar gases. As reviewed in \ref{subsec:Vk} and~\ref{subsec:BMF}, the mode driving the instability can be either phononic or rotonic depending on the relative configuration of trap, excitation, and dipoles, see Fig.~\ref{fig:dispersion}. This underlies the quantum stabilization of diverse ground states, being either unmodulated or modulated. 

\subsubsection{Phonon-driven MF instabilities and unmodulated states}\label{par:phonon_instab}

Phonon-driven MF instabilities underlie quantum-stabilized unmodulated ground states~\footnote{We note that phonon instabilities can lead to modulated dynamical states as observed in contact interacting gases, see e.g.~\cite{Nguyen2017fom}}. These states have yet unique properties: Even in the absence of confinement in some or all spatial direction(s), they are (non-uniform) localized solutions~\cite{Baillie2016sbd,Baillie2017ceo,Wachtler2016gsp}.  They are thus self-bound under the competition of MF and beyond-MF interactions, and form an ultradilute liquid, warranting their droplet designation~\cite{FerrierBarbut2018qlg}. In trapped systems, droplet states can also occur and display the distinct property of not expanding upon release of the trapping potential $V(\br)$~\cite{schmitt2016sbd,Bisset2016gsp,Wachtler2016gsp}. 

Beyond their defining self-binding character, droplet states exhibit other outstanding features: They have an unusual shape, being highly elongated along the dipole direction. 
Droplets containing a sufficiently large number of atoms, $N$, present a broad central region of uniform density, and increasing $N$ further does not raise the central density but instead causes the droplet to extend along the dipole orientation~\cite{Baillie2016sbd}. This underlies the state's weak compressibility and has significant implications for its collective excitations~\cite{Chomaz2016qfd,Wachtler2016gsp,Baillie2017ceo,Pal2022}, see later discussion in \ref{subsec:exc_droplets}. These features are likewise reminiscent of a liquid behavior.

Soon after the first observation of quantum stabilization in Ref.~\cite{Kadau2016otr} (see \ref{subsec:BMF}), the investigation of unmodulated droplets became a focus of interest. The motivation is twofold: (i) in a fully trapped system, the transition from MF-stable BEC to droplet occurs through a smooth crossover~\cite{Bisset2016gsp,Wachtler2016gsp}, thereby limiting the introduction of excitations into the system, and (ii) the quantitative study of the droplet properties allows to test that quantum fluctuations drive the stabilization~\cite{FerrierBarbut2016ooq, Chomaz2016qfd, schmitt2016sbd, FerrierBarbut2018ooa}.  

\subsubsection{Roton-driven MF instabilities and modulated states}\label{par:roton_instab}
Roton-driven MF instabilities instead underlie the formation of crystallized states, i.e.~states with a spontaneous modulation in their density, as previously suggested for superfluid helium and discussed in \ref{subsec:Vk},~\ref{subsec:BMF}~\cite{Nozieres2004itr,Feynman1954}. Two types of quantum-stabilized modulated states should be distinguished. In one case, the transition from an unmodulated superfluid to a modulated state through roton softening may allow the modulated state to retain its global superfluidity, forming a supersolid~\cite{Gross1957uto, Giovanazzi2002dmo,Santos2003rms,Boninsegni2012csw,Yukalov2020, Chomaz2023dipolar}. Supersolids are thus a special case of modulated states combining seemingly antithetical properties -- those of a superfluid, where atoms are delocalized, and those of a solid, where atoms tend to be localized at fixed positions of space. That underlies the spontaneous breaking of two continuous symmetries, translational and gauge. In a second scenario, the modulated state does not retain the initial superfluidity and instead forms an insulating crystal. In this case, the atoms are delocalized only on one crystal site and are prevented from flowing between the different sites. 

The superfluid character of a modulated state intimately connects to the contrast $\mathcal{C}>0$ of its density modulation ($\mathcal{C}=0$ is the uniform state). 
For a low-enough $\mathcal{C}$, the density peaks associated with the crystal sites have significant overlap. This enables the atoms to tunnel from one site to the next, and global superfluidity is preserved. Conversely, a high $\mathcal{C}$ indicates a reduced overlap between peaks and yields a low tunneling rate, insufficient to maintain global superfluidity. The superfluid fraction of modulated states is always reduced compared to unity, and its value can be quantitatively related to the density patterns through the Leggett bounds~\cite{Leggett1970cas,Sepulveda2008nri}. We note that the distinction between supersolid and insulating solid is not captured by a standard eGPE treatment, which inherently assumes phase coherence (therefore global superfluidity). Nevertheless, an empirical threshold on the contrast is typically used to distinguish between these two situations, see e.g.~\cite{Blakie2020sia}.

Returning to our first point: roton softening and quantum-stabilized modulated states are interlinked. As introduced in \ref{par:excitation_rotons}, the occurrence of a roton requires stronger confinement along the dipole direction than in other direction(s), along which instead the roton-maxon dispersion arises, see Fig.~\ref{fig:dispersion}(f). The trap geometry, particularly the number of tightly confined directions, strongly influences the structure of both the roton modes and the underlying stabilized modulated states. In tubular geometries, roton modes develop along the single weakly confined direction, and the underlying modulated states acquire a linear crystal structure along this axis~\cite{Chomaz2018oot, Ancilotto2019sbo,Tanzi2019ooa,Chomaz2019lla,Bottcher2019tsp}. In pancake geometries, where two spatial directions are weakly confined, roton modes emerge in both these directions, allowing for more complex two-dimensional crystal arrangements of the stabilized states~\cite{Lu2015a,Zhang2019saa,Ripley2023Two}. The resulting crystal structure further depends on factors such as density, interaction strength, and confinement. 

As discussed in \ref{par:fulltrapexc}, in a full 3D trap, the nature of mode driving the instability may be altered. Nonetheless, if the instability arises from non-phononic modes, density-modulated ground states may still form. This is in particular the case when the instability stems from the softening of modes with high angular momentum, known as angular rotons~\cite{Ronen2007raa,Martin2012a, Chomaz2023dipolar, Schmidt2021rei, Hertkorn2021sit}. 

The remainder of this section reviews the wealth of modulated ground states observed and predicted depending on the trap and gas parameters.

\begin{figure*}[ht!]
\centering
\includegraphics[width=1.\textwidth]{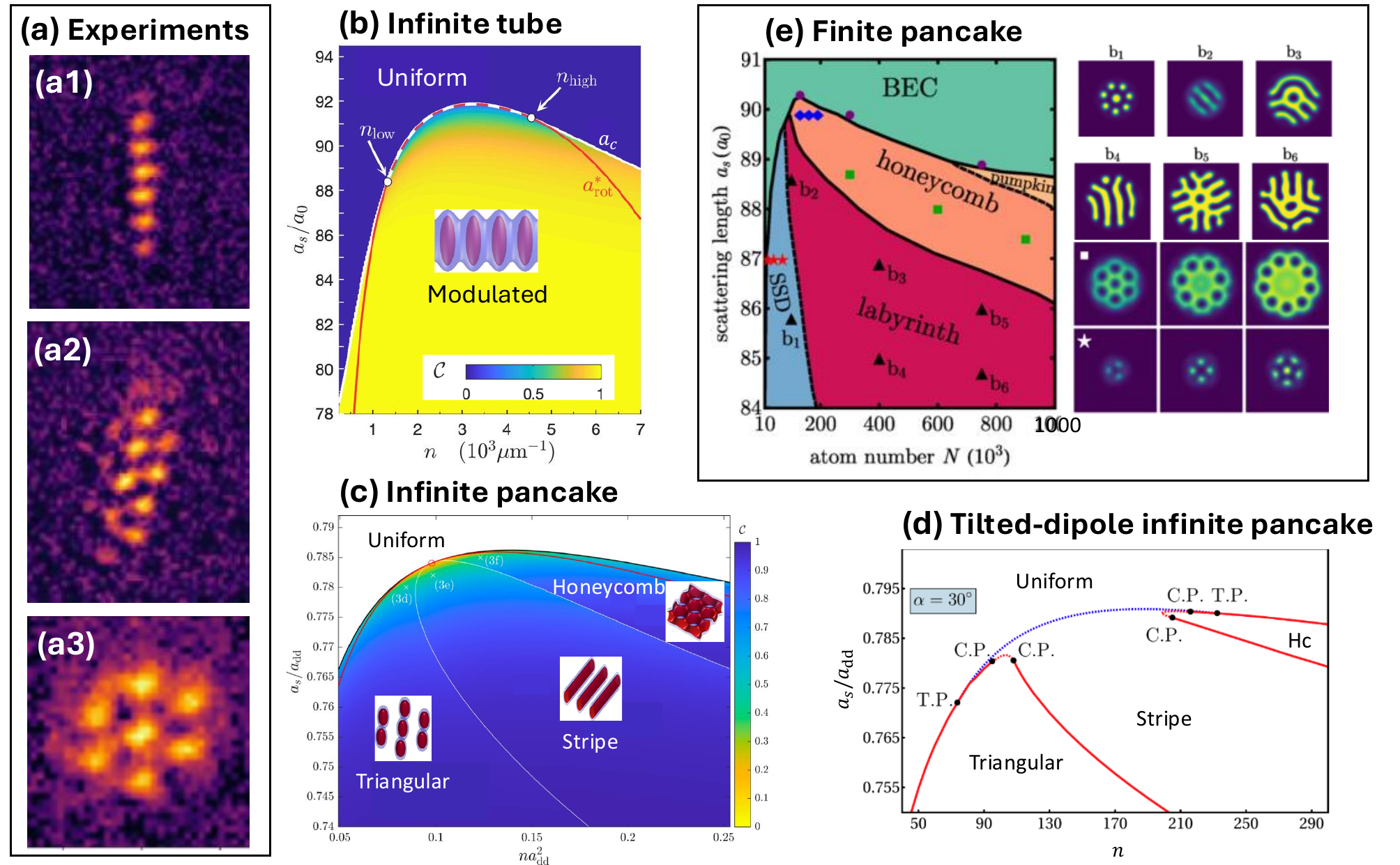}

\caption{\textbf{Quantum-stabilized modulated ground states}. 
(a) Examples of measured density distributions of quantum-stabilized states of \Dyb~in different trap geometries: (a1) cigar-shaped trap, (a2) surfboard trap, (a3) circular pancake trap, taken from~\cite{Norcia2021tds,Bland2021tds}. (b, c) Theoretical phase diagrams of \Dyb~gases with transverse dipole orientation in infinite cigar and pancake traps, adapted from~\cite{Blakie2020sia} and \cite{Ripley2023Two} respectively. The color map shows the modulation contrast $\mathcal{C}$. The region with $\mathcal{C}=0$ is the uniform state. The roton excitation of this state fully softens at $\accrot$ (red line). The transition to the modulated states occurs at $\ac$ (white and black lines, respectively). The modulated states are illustrated via density isosurfaces and the transitions between different crystal structures identified (white lines). (d) Same as (c), but with dipoles tilted by $\alpha=30^{\textrm{o}}$ compared to the pancake confinement axis, taken from~\cite{Lima2025sdp}. Continuous (resp.~discontinuous) transitions are indicated by dotted (resp.~solid) lines with end points identified by C.P. or T.P. (e) Phase diagram of \Dyb~gases in a pancake trap with radial and axial frequencies $\omega_{\parallel,\perp}=2\pi\times(125,250)$\,Hz, taken from \cite{Hertkorn2021pfi}. Transitions are marked by black lines. Different crystal structures are identified, and their density profiles (markers) illustrated in subpanels.
}
\label{fig:PD}
\end{figure*} 

\subsection{Experimental observations of modulated states}\label{sec:PDexp}

 The first experimental observations of quantum-stabilized states were realized in pancake traps with transverse dipole orientation and displayed crystal structures~\cite{Kadau2016otr,FerrierBarbut2016ooq,Wenzel2017ssi}, see \ref{subsec:BMF}. However, these were highly excited states, while the ground state was unmodulated due to the low atom number and shallow trap employed~\cite{Wachtler2016qfi}, see also later discussion in \ref{subsec:trappedPD}. Later experiments realized multi-droplet states close to the ground-state configuration using tighter pancake traps, but no global phase coherence was observed in such settings~\cite{Wenzel2017ssi}. Refs.~\cite{Kadau2016otr} (\cite{Wenzel2017ssi}) used pancake traps with a radial frequency of approximately 50\,Hz, axial frequencies of 133\,Hz (varying from 250\,Hz to 1700\,Hz), and $15\,000$ ($5\,000$) atoms.

The search for globally phase-coherent crystal ground states was revived following the work of Ref.~\cite{Chomaz2018oot}, which demonstrated the coherent population of the roton mode after an interaction quench in a weakly confined cigar-shaped gas of \Erb. In 2019, using a similar trapping potential but smaller and slower interaction ramps, three groups observed the formation of metastable states with supersolid properties using either \Erb, \Dyb, or \Dybb~\cite{Tanzi2019ooa,Chomaz2019lla,Bottcher2019tsp}. For \Dyb, supersolid states were also formed via direct evaporative cooling from a thermal state at the final interaction settings~\cite{Chomaz2019lla,Sohmen2021bla}.  These experiments used relatively shallow cigar traps with axial frequencies between 16 and 32\,Hz and averaged radial frequencies between 65 and 260\,Hz. The dipoles were aligned along one radial direction, and between 40\,000 and 80\,000 atoms were trapped. The supersolids showed linear crystal structures with a few crystal sites, each populated by several thousand atoms, and thereby called droplets, see Fig.\,\ref{fig:PD}(a). These states were observed to live from a few tens of ms for \Dybb~and \Erb~to a few hundreds of ms for \Dyb~ after interaction ramps. Longer lifetimes, on the order of seconds, were achieved using temperature ramps~\cite{Sohmen2021bla}.

Following these initial observations, the search for more complex crystal structures drove particular interest. Supersolids with nonlinear crystal structures were successfully realized, with first zigzag supersolids in surfboard-shaped traps~\cite{Norcia2021tds}, see Fig.\,\ref{fig:PD}(a2), and then circularly symmetric triangular supersolids (hexagon supersolids)  in a round pancake trap~\cite{Bland2021tds}, see Fig.\,\ref{fig:PD}(a3). These supersolids were produced via temperature ramps, and exhibit second-long lifetimes. Compared to the earlier experiments of Refs.~\cite{Kadau2016otr, Wenzel2017ssi}, significantly larger atom numbers were used with N=$65\,000$ and $44\,000$, while the traps remained relatively shallow with frequencies (33,75,167)\,Hz and (47,44,133)\,Hz in the three spatial directions, for Ref.~\cite{Norcia2021tds} and~\cite{Bland2021tds} respectively. In both cases, the dipoles were aligned along the tightest trap direction.

\subsection{Theoretical predictions of modulated ground states in infinite geometries}\label{subsec:infinitePD}

Quantum-stabilized modulated states require at least one confined axis, see \ref{par:roton_instab}. Phase diagrams can be theoretically calculated by considering partly infinite systems, analogous to the approach of \ref{subsec:confexc} for the excitation spectrum. Both infinite cigar (one unconfined, two confined directions)~\cite{Ancilotto2019sbo, Blakie2020sia,Ilg2023gss,Smith2023supersolidity,Blakie2023}, and pancake geometries (two unconfined, one confined direction)~\cite{Zhang2019saa,Ripley2023Two,Lima2025sdp,Zhang2024,Zhang2023vat} can be examined. Periodic boundary conditions in the unconfined directions ensure that the system is effectively infinite along these axes. The ground-state phase diagram can be computed using the eGPE introduced in \ref{subsec:BMF}, provided that the system size (or equivalently the crystal unit cell) is also adjusted and the energy minimized with respect to this parameter.

These systems can also be simulated by other methods, in particular quantum Monte Carlo (QMC) 
approaches~\cite{Saito2016pim,Macia2016dot,Cinti2017caq,Bombin2017dbs,kora2019patterned,Bombin2019bkt,Kora2020dbi,Staudinger2023sdl,Boettcher2019ddq}. QMC methods are exact and account for fluctuation effects non-perturbatively. However, they are numerically intensive, limiting so far their exploration to relatively small $N$. Therefore, reaching density-modulated states in QMC simulations typically requires higher mean densities than those used in experiments. In this regime, QMC simulations have confirmed the existence of supersolid states, and the crystal structures identified are analogous to those predicted with the eGPE and discussed below~\cite{Cinti2017caq,Bombin2017dbs,kora2019patterned,Bombin2019bkt,Kora2020dbi,Staudinger2023sdl}.

Figures \ref{fig:PD}(d-f) present eGPE phase diagrams as a function of $\as$ and $n$ for different infinite-geometry configurations. In all cases, the ground state transitions from a uniform MF-stable superfluid for $\as\geq\ac(n)$ to a quantum-stabilized modulated state for $\as<\ac(n)$ with $\ac(n)$ being the density-dependent critical scattering length. 

\subsubsection{Tube geometries}\label{par:tubesupersolids}
In a tubular trap, the crystal structure is always linear, but the nature of the transition changes depending on $n$~\cite{Blakie2020sia}. The transition is predicted to be discontinuous at low and high densities, but continuous at intermediate densities, see Fig.\,\ref{fig:PD}(b). 

In the intermediate density regime, the continuity of the transition implies that the density modulation contrast evolves continuously from $\mathcal{C}=0$ above the transition ($\as \geq \ac$) to $\mathcal{C}>0$ below it ($\as < \ac$). Therefore, a supersolid ground state exists within an intermediate range just below the transition ($\as\lesssim \ac$). The width in $\as$  of this supersolid region varies with $n$ and narrows towards the intermediate-density range edges, see Fig.\,\ref{fig:PD}(b). The transition here occurs as the roton mode fully softens, meaning $\ac=\accrot$. Thereby, the emerging modulated state directly connects to the softened roton mode and inherits its structure. Discontinuities in the first derivative of the dispersion relation occur at the transition, indicating its second-order character, see later discussion in \ref{subsec:excitations}~\cite{Blakie2023,Ilg2023gss}. As  $\as$ decreases further below $\ac$, $\mathcal{C}$ increases until the system transitions into an insulating droplet array.

Beyond the intermediate-density-range lower edge, in the low-density regime, the system transitions directly to an array of well-isolated droplets with $\mathcal{C}\sim 1$ and no supersolid can form, see Fig.\,\ref{fig:PD}(b). This happens because quantum stabilization, i.e.~MF attraction being overcome by the LHY repulsion \eqref{eq:LHY}, requires locally high densities. These densities are here achieved by concentrating many atoms in the center of the droplets at the expense of their overlap. 

Beyond the intermediate-density-range upper edge, in the high-density regime, instead, a supersolid ground state (i.e.~a modulated state with $\mathcal{C}< 1$)  can exist just below the transition despite being discontinuous, see Fig.\,\ref{fig:PD}(b). The structure of this state differs from the roton mode of the MF-stable superfluid, and the transition involves a rearrangement of the transverse density profile.

For the first-order transitions encountered at high and low densities, bistability occurs, with the ground state on one side of the transition remaining a metastable state on the other side. In particular, the uniform state persists as a metastable state on top of the modulated ground state for $\as<\ac$ (white line in Fig.\,\ref{fig:PD}(b)) down to $\as=\accrot$, when its roton mode fully softens (red line in Fig.\,\ref{fig:PD}(b)).

The intermediate density regime is particularly favorable for creating supersolid states via an interaction ramp as the transition continuity minimizes excitations during its crossing. Furthermore, the critical scattering length $\ac$ increases with $n$ in the low-density regime, decreases in the high-density regime, and the intermediate-density regime develops around the maximum of $\ac(n)$, which is likewise favorable for experiments, see also \ref{subsec:BMF}.  Refs.~\cite{Tanzi2019ooa, Chomaz2019lla, Bottcher2019tsp} realized this intermediate regime, see also Refs.~\cite{Blakie2020sia, Hertkorn2019fot, Hertkorn2021dfa,Biagioni2022dci} and \ref{sec:PDexp}. 

\subsubsection{Pancake geometries}\label{par:pancakesupersolids}
In pancake traps, not only the order of the uniform-to-modulated phase transition but also the structure of the modulated states themselves varies with $n$ and $\edd$, see Fig.\,\ref{fig:PD}(c). This, in turn, implies the existence of additional \emph{structural} transitions between modulated ground states of different crystal configurations. Density-modulated ground states may adopt three different configurations illustrated in Fig.\,\ref{fig:PD}(c): triangular (also sometimes called hexagonal) at low densities, stripe at intermediate densities, and honeycomb at high densities~\cite{Ripley2023Two,Lima2025sdp,Zhang2023vat}.

Most transitions in the diagram of Fig.\,\ref{fig:PD}(c) are predicted to be of first order, except at the tricritical point where the transition is of second order and the uniform state may transition directly to a stripe state~\cite{Zhang2019saa,Ripley2023Two,Lima2025sdp}. 
Despite the discontinuous nature of these transitions, all three crystal structures can host supersolid states, i.e.~have significant overlap between density peaks with $\mathcal{C}<1$ so that global superfluidity is preserved. 

Similar to the tube geometry, the first-order transitions underlie bistability, with the ground state remaining as a metastable state on the other side of the transition. This holds for the uniform-to-modulated transition with the uniform state being a metastable state down to the roton instability like in \ref{par:tubesupersolids} (see red line Fig.\,\ref{fig:PD}(c)), but also for the structural transitions between stripes and the other crystal structures, where the different modulated states remain metastable down to a point where one of their sound modes becomes unstable~\cite{Blakie2025dpa}, see also later discussion in \ref{subsec:exc_supersolid}.

So far, experimental realizations of 2D-crystal supersolids have been made at relatively low densities and trapping strength, and thereby limited to triangular phases~\cite{Norcia2021tds,Bland2021tds}, see Fig.\,\ref{fig:PD}(a2,a3). These states have been created by temperature instead of interaction ramps, which avoids crossing discontinuous transitions, and thereby minimizes excitations~\cite{Bland2021tds}, see \ref{sec:PDexp}.

\subsubsection{Effect of dipole tilt}
 The phase diagrams of Fig.\,\ref{fig:PD}(b,c) are shown for dipoles oriented along a tightly confined direction.  Tilting the dipoles by an angle $\alpha$ with respect to this direction and towards an unconfined direction provides an additional tuning parameter, controllable in experiments by adjusting the orientation of the applied bias field $\vB$ using several pairs of coils. In the extreme case of $\alpha=\pi/2$, one recovers the situation described at the beginning of \ref{subsec:modulation}, where phonon softening drives the MF instability and the quantum-stabilized ground state is a single-droplet state. Instead, for sufficiently small tilt angles, density-modulated ground states occur, yet the phase diagram is modified compared to Fig.\,\ref{fig:PD}(b,c). 

In tube geometries, the crystal structures remain linear, but both the critical scattering length $\ac$ and the crystal spacing increase with $\alpha$.  For pancake geometries, an example phase diagram with  $\alpha=\pi/6$ is shown in Fig.\,\ref{fig:PD}(d)~\cite{Lima2025sdp}.  While this diagram features the same phases and their same density ordering as the $\alpha=0$ diagram of Fig.\,\ref{fig:PD}(c), the stripe phase is favored at the expense of the triangular and honeycomb phases~\cite{Lima2025sdp,Wenzel2017ssi}. The dipole tilt also influences the orientation of the stripes. While for $\alpha=0$ this orientation is random, for $\alpha>0$, the stripes orient themselves along the unconfined direction in which the dipoles are tilted. 

In addition to modifying the phases' extents, the dipole tilt further impacts the phase transitions and their nature. In pancake geometries, it extends the density range for which a direct transition between uniform and stripe state exists, transforming the tricritical point of Fig.\,\ref{fig:PD}(c) into a line in Fig.\,\ref{fig:PD}(d) (blue dotted line). Along this line, the uniform-to-modulated transition is predicted to be continuous~\cite{Lima2025sdp}. Similarly, the dipole tilt introduces regimes where the structural transitions between stripe and triangular or honeycomb phases become continuous (red dotted lines in Fig.\,\ref{fig:PD}(d)). This configuration of experimental interest has so far only been explored in early experiments~\cite{Wenzel2017ssi} where supersolidity was not achieved, see \ref{sec:PDexp}.

\subsection{Theoretical predictions of modulated states in trapped systems}\label{subsec:trappedPD}

Beyond the case of infinite geometries, the phase diagram of dipolar gases in fully finite traps has attracted considerable interest. Studies have explored various trap geometries ranging from finite tubular~\cite{poli2021msi,Chomaz2019lla,Bottcher2019tsp,Tanzi2019ooa,Biagioni2022dci} to pancake-like~\cite{Wenzel2017ssi, Baillie2018,Zhang2021pos, Hertkorn2021pfi,Hertkorn2021sit} passing by fully anisotropic ones~\cite{Norcia2021tds}. This interest stems not just from the experimental relevance of such configurations, but also from the unique effects they induce, including beyond quantitative shifts, qualitative differences, and unprecedented crystal structures in the system's ground state.  
The traps used in these theoretical works are typically harmonic, approximating the experimental configurations using Gaussian beam traps. Such traps are not only finite but also inhomogeneous, resulting in a distinctive interplay with the DDI. 

In finite traps, the atom number $N$ should be considered as a control parameter instead of the mean density $n$. Considering $N$ introduces finite-size effects, which, combined with the trap frequencies, limit the number of crystal sites that the ground state hosts. 
For too small $N$, the number of crystal sites is reduced to one, and unmodulated ground states are found in trap configurations for which larger $N$ or infinite-system phase diagrams display density modulation, see e.g.~\cite{Chomaz2019lla,Wachtler2016qfi}. This was the case in the early experiment of Ref.~\cite{Kadau2016otr}, see \ref{sec:PDexp}.

Typical phase diagrams vary $N$ and $\edd$ for fixed trap parameters. Fig.\,\ref{fig:PD}(e) shows an example for a pancake geometry with transverse polarization and moderate ratio $\omega_\perp/\omega_{\parallel} = 2$ between axial and radial trap frequencies~\cite{Hertkorn2021pfi}. For this moderate aspect ratio, instead of a conventional roton mode, an angular roton drives the MF instability, see \ref{subsec:modulation}~\cite{Hertkorn2021sit}. Therefore this scenario notably deviates from the considerations of \ref{subsec:infinitePD}. Nonetheless, Fig.\,\ref{fig:PD}(e) shows modulated states with familiar patterns -- triangular, stripe, honeycomb -- appearing in same sequence with increasing $N$ as with increasing $n$ in infinite systems (see Fig.\,\ref{fig:PD}(c)). This demonstrates the relevance of infinite-system predictions for understanding finite systems, even with small aspect ratios.  

Besides these familiar phases, Fig.\,\ref{fig:PD}(e) features modulated states with novel structures -- pumpkin or labyrinth -- illustrated in the subpanels. 
The labyrinth phase is particularly fascinating as the underlying crystal structures lack symmetries, and many configurations are almost degenerate. These characteristics being reminiscent of glassy behaviors, this phase was dubbed \emph{superglass} in Ref.~\cite{Hertkorn2021pfi}. It occurs over a wide range of parameters, partly or fully replacing the stripe phase, and potentially emerges from frustration between the in-plane confinement and the random orientation of the stripes predicted in infinite pancakes.  

Finite systems with larger trap aspect ratios have also been investigated, in particular in Ref.~\cite{Lu2015sds} published before the quantum-stabilization discovery, and therefore based on a different stabilization mechanism. Nonetheless, its infinite-system phase diagram closely resembles Fig.\,\ref{fig:PD}(c). Fully trapped systems were here studied from a local density approximation standpoint: the gas is treated as being locally uniformly trapped, but with a chemical potential (or average density) that varies spatially. This links the crystal structure at different trap positions to different points in the infinite-system phase diagram, changing from honeycomb to stripe to triangular when moving from the trap's inner to outer regions. In the smaller aspect ratio trap of Fig.\,\ref{fig:PD}(e), some crystal structures may be interpreted using this analysis, e.g.~the square markers transition from uniform to honeycomb from inner to outer regions. Analogously, low-density uniform halos have been predicted in the trap outer regions around linear or triangular supersolids occurring at the center of tube or pancake traps~\cite{Wenzel2017ssi,Baillie2018,Chomaz2019lla,Bottcher2019tsp,Tanzi2019ooa}.

\subsection{Effect of temperature}\label{subsec:temp} 

The phase diagrams of Fig.\,\ref{fig:PD}(b-e) are calculated in the zero-temperature limit. However, in experiments, the temperature, $T$, is inherently finite and can be tuned. This parameter is expected to influence the phase behavior in two main ways: it controls the emergence of the quantum phases themselves;  it introduces thermal fluctuations, which add to but differ from the quantum fluctuations, and thus modify the stabilized phase diagram at low but finite $T$. 

As reported in \ref{sec:PDexp}, experiments have shown that the gas can transition from a thermal state to a supersolid by tuning $T$~\cite{Chomaz2019lla, Sohmen2021bla, Norcia2021tds}. During this transition, two symmetries are spontaneously broken: translational and gauge, linked to solid and superfluid orders, respectively. These two symmetries can be broken simultaneously or successively when varying $T$. The nature, number, and order of the thermally driven phase transitions are subjects of active investigation. In experiments involving evaporative cooling in cigar-shaped traps, the gas has been observed to first transition from a thermal state to a normal solid state with local phase coherence, before achieving global phase coherence and forming a supersolid as $T$ decreases~\cite{Sohmen2021bla}.  This suggests that translational symmetry can break at higher $T$ than gauge symmetry. However, since not only temperature but also total atom number and trap geometry evolve during the evaporative cooling process, the generality of these observations may be limited. Experiments in cigar-shaped traps have also indicated that density-modulated states are favored over uniform ones as $T$ increases at a fixed condensate atom number, revealing influences of $T$ on the phase diagram~\cite{Sohmen2021bla,SanchezBaena2023had}. So far, experimental studies on thermal effects and transitions in pancake geometries have been limited~\cite{Norcia2021tds, Bland2021tds}.  

Theoretically, the finite-$T$ regime is not directly accessible in the extended MF theory of \ref{subsec:BMF} based on the eGPE. However, it can be partly captured either by making refinements to the eGPE~\cite{Bland2021tds,SanchezBaena2024ssp,SanchezBaena2023had,He2025aet} or by starting from a different standpoint and using QMC treatments, which are exact but computationally limited, see \ref{subsec:infinitePD}~\cite{Bombin2019bkt}.  

Refinements of the eGPE have followed different approaches: (i) A stochastic projected version of the eGPE adds two terms set by empirical parameters to the standard eGPE: a dissipative term that models the coupling of the classical field $\psi$ to high-energy modes, and a dynamical noise term that embodies thermal fluctuations~\cite{Bland2021tds}. This approach enables simulations of the dynamics of $\psi$ in the presence of a thermal bath. (ii) An eGPE version including thermal fluctuation corrections adds, alongside with $\mu_{\rm LHY}$ from the standard eGPE, a thermal chemical potential term, $\mu_{\rm th}$, to the right-hand bracket of \eqref{eq:GPE}~\cite{SanchezBaena2024ssp,SanchezBaena2023had,He2025aet}. Akin to $\mu_{\rm LHY}$, $\mu_{\rm th}$ accounts for the energy shift of $\psi$ induced by its interactions with thermal excitations. It is derived assuming local density approximation and a thermal population of the Bogoliubov excitation spectrum \eqref{eq:Bdg} of a uniform gas (see \ref{subsec:unconfexc}).

Based on the stochastic projected eGPE (i), evaporative cooling ramps in which both temperature $T$ and chemical potential $\mu_0$ vary have been simulated~\cite{Bland2021tds}. These simulations confirm the experimental observation of a solid order appearing at higher $T$ than superfluid order, here in the case of pancake traps. They also show that supersolid states are stable against thermal fluctuations. Similar results have been obtained using QMC simulations, which further predict that the underlying transitions follow the Berezinskii-Kosterlitz-Thouless (BKT) mechanism~\cite{Bombin2019bkt}.

Based on the thermally extended eGPE (ii), phase diagrams including $T$ as a control parameter have been calculated both in tubular ~\cite{SanchezBaena2023had} and pancake-shaped geometries~\cite{He2025aet} with transverse polarization. These confirm and generalize the experimental observation that supersolid states are favored over uniform states when $T$ increases at a fixed condensate atom number (i.e.~$\int d^3r|\psi(\br)|^2$) and density (i.e.~$|\psi(\br)|^2$) in finite and infinite geometries, respectively. This occurs as $\mu_{\rm th}$ also contributes in the dispersion relation \eqref{eq:Bdg} and makes the roton soften at lower $\edd$ as $T$ increases. The thermally extended eGPE (ii) does not directly account for incoherent effects, making the thermal transition inaccessible via pure ground-state calculations in this formalism (analogous to the supersolid-to-insulated-crystal transition in the $T=0$ simulations, see \ref{subsec:modulation}). 
Phase diagrams for varying condensate density and interaction parameters at fixed $T$ calculated using the thermally extended eGPE (ii) in both tubular ~\cite{SanchezBaena2024ssp} and pancake geometries~\cite{He2025aet} with transverse polarization show that higher $T$ slightly increases the critical $\ac$ for a fixed condensate density. In tubular geometries, the intermediate density range supporting a continuous transition (see \ref{par:tubesupersolids}) slightly shifts to lower $n$. In pancake geometries, crystal structures occur with the same density ordering as in the $T=0$ diagram (see \ref{par:pancakesupersolids}) but their density ranges are shifted down. The shift is large for shallow transverse traps with $\ell_\perp\sim 100\add$~\cite{He2025aet}, and decreases with increasing confinement tightness (decreasing $\ell_\perp$) as the thermal fluctuations lessen in strength with increasing 3D density. 

\begin{figure*}[ht!]
\centering
\includegraphics[width=1.\textwidth]{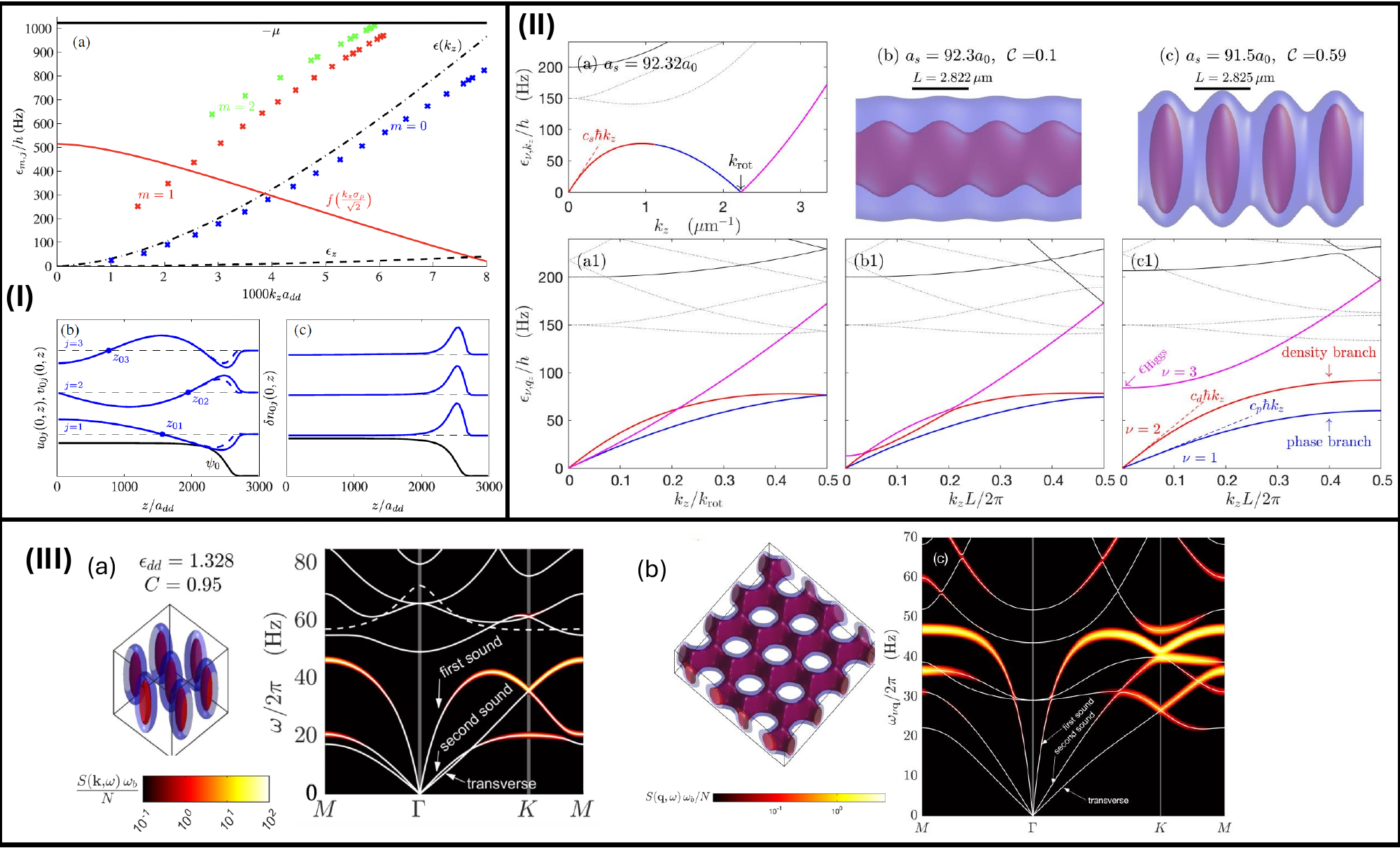}

\caption{{\textbf{Excitation spectra of quantum-stabilized states}. (I) Spectrum of an unconfined droplet, taken from~\cite{Baillie2017ceo}: (I.a) Bogoliubov excitation energies with three branches $m=0,1,2$ (blue, red, green crosses). The lines show characteristic energies: kinetic energy (dashed), dispersion relation \eqref{eq:Bdg} with LHY correction (dashed-dotted), $\Vdd$ (solid red). Examples of (I.b) quasiparticle amplitudes and corresponding density fluctuations (I.c) (blue) compared to the ground state density (black) for the lowest modes of the $m=0$ branch. The positions $z_{0,i}$ are used to assign the momenta in (I.a). (II) Dispersion of a $D=1$ supersolid, taken from~\cite{Blakie2023}: (II.a,a1) at $\as=\ac$, illustrating the Brillouin-zone mapping. (II.b1-c1) at lower $\as$, showing the evolution of the different branches. The upper subpanels (b-c) show the ground state structures.  (III) Spectrum of the $D=2$ supersolids with illustrated crystal structures on the left, for (III.a) a triangular supersolid, taken from~\cite{Poli2024eoa}, (III.b) a honeycomb supersolid, taken from~\cite{Blakie2025dpa}. The dispersion relation is illustrated by the dynamic structure factor (color map), and the Bogoliubov excitation energies (white lines). 
}} 
\label{fig:spectra}
\end{figure*}

\section{Elementary excitations of the quantum-stabilized states}\label{subsec:excitations}


The unique properties of the quantum-stabilized states manifest in their dispersion relation and elementary excitation spectrum. This applies to both single-droplet states and supersolids, though with markedly different features. Analogously to the case of uniform states discussed in \ref{subsec:Vk}, the excitation features of quantum-stabilized states relate to a momentum-dependent interaction potential, $\tVint(\vk)$, that stems from the competition of the long-range anisotropic DDI with the state geometry.
This competition is renewed compared to uniform states due to the novel morphologies adopted by quantum-stabilized states, see \ref{sec:PD}. 

The dispersion relation \eqref{eq:Bdg} does not directly apply to quantum-stabilized states because of their non-uniform character.  Instead, their dispersion and the underlying elementary excitations can be derived within Bogoliubov theory introduced in \ref{par:fulltrapexc} for fully trapped systems. This approach consists in linearizing the time-dependent GPE \eqref{eq:GPE} around the ground-state wavefunction $\psi$, while applying perturbations to $\psi$, called quasiparticles. 
This applies to quantum-stabilized states provided that the GPE is extended by the quantum-fluctuation term discussed in \ref{subsec:BMF} and possibly by the thermal-fluctuation term discussed in \ref{subsec:temp}. Hereafter, we review the excitation features of single-droplet states and supersolids based on the Bogoliubov approach.
\subsection{Excitation Spectrum of Droplets}\label{subsec:exc_droplets}

As reviewed in \ref{subsec:modulation}, single-droplet states form in regimes where phonon modes soften and drive the MF instability. This occurs in geometries where the dipoles are oriented along a weakly confined or unconfined direction. The resulting droplet states, stabilized beyond the MF instability, are finite, localized, and highly stretched in the dipole direction, denoted $z$. Droplets appear both in trapped and fully unconfined settings. Droplets of large enough $N$ reach the incompressible regime. 

Due to their finite character, droplets exhibit a discrete excitation spectrum, with mode spacing dictated by the droplets' sizes. In fully unconfined settings, mode energies are limited by the droplet's chemical potential, thereby setting a limit to the number of modes, which increases with $N$.
The low-energy excitation spectrum is dominated by excitations along $z$, as this is the droplet's largest dimension~\cite{Baillie2017ceo,Wachtler2016gsp}. This corresponds to the excitation configuration modeled by Eq.\,\eqref{eq:Bdg} in \ref{par:excitation_phonons} for the uniform state.

Due to the droplet's non-uniform character, momentum is not a good quantum number, and the dispersion relation~\eqref{eq:Bdg} does not directly apply. However, the lowest excitation branch of sufficiently large droplets proves to connect with \eqref{eq:Bdg} through a mode-momentum assignment and the inclusion of a term accounting for the LHY stabilization, see Fig.\,\ref{fig:spectra} (I.a)~\cite{Baillie2017ceo}. The former is achieved by considering the first spatial cancellation of the quasiparticle amplitudes, see Fig.\,\ref{fig:spectra} (I.b). The latter is done by adding $\frac{3}{5}\gamma_{\rm QF}n^{1/2}$ [see Eq.\,\eqref{eq:LHY}] to $\Vint(\vk)$ in \eqref{eq:Bdg}~\cite{Pal2022,Baillie2017ceo}. This correction is also relevant for uniform superfluids near the phonon instability, resulting in a lowering of the instability threshold $\ac$. 

The droplet state formed at $\as<\ac$  exhibits a higher density than the superfluid state, and a transverse size governed by its self-bound nature rather than the trapping potential.  The density increase, combined with the added LHY term in \eqref{eq:Bdg}, and the transverse rearrangement lead to a hardening of the lowest-energy modes, thus preventing instability. In trapped systems, the excitation gap remains finite and continuous as $\as$ varies, indicating a smooth crossover from uniform superfluid to droplet~\cite{Wachtler2016gsp, Baillie2017ceo}. In fully untrapped systems, the lowest droplet mode goes soft at the transition towards a uniform state~\cite{Baillie2017ceo}.

For droplets in the incompressible regime, the dispersion relation does not exhibit a linear phononic behavior. However, unlike non-interacting BECs where a similar deviation arises from dominant kinetic effects, the droplet nonlinear dispersion stems from interaction effects, namely the $k$-dependence of $\Vint$~\cite{Baillie2017ceo}, see Fig.\,\ref{fig:spectra} (I.a). The modes of incompressible droplets also have distinct features: they show minimal bulk density fluctuations and dominant surface ones, the lowest excitation being of quadrupolar character~\cite{Wachtler2016gsp,Baillie2017ceo}, see Fig.\,\ref{fig:spectra} (I.c).

\subsection{Excitation Spectrum of Supersolids}\label{subsec:exc_supersolid}

For roton-driven MF instability, quantum-stabilized density-modulated ground states can form. Their dispersion relation intimately relates to and reflects the crystal structure. For states with periodic density modulation, such as those formed in partly infinite geometries (see \ref{subsec:infinitePD}), the Bogoliubov quasiparticle amplitudes are well described by Bloch wavefunctions. These are characterized by a quasimomentum $q$ defined on the first Brillouin zone $q \in ]-\pi/L,\pi/L]$ with $L$ the supersolid periodicity and a branch index $\nu$~\cite{Roccuzzo2019,Sindik2023sound,Blakie2023,Ilg2023gss,Poli2024eoa,Blakie2025dpa}. 

Supersolids are density-modulated states that also manifest global superfluidity. A distinctive feature of the supersolid excitation spectrum is that it features multiple gapless excitation branches at $q\rightarrow 0$. According to the Nambu-Goldstone theorem, the number of gapless branches reflects the number of spontaneously broken continuous symmetries~\cite{Watanabe2012}. For supersolids with a crystal of dimension $D$, there are $D+1$ gapless branches, corresponding to $D$ broken translation symmetries and 1 broken gauge symmetry. The supersolid excitation spectrum also presents gapped modes, a notable one being the Higgs mode, associated with amplitude fluctuations of the superfluid order parameter~\cite{Pekker2015ami}. 

\subsubsection{Tube geometries}
In tubular traps, $D=1$ and the supersolids exhibit two Goldstone modes~\cite{Natale2019eso,Tanzi2019ooa,Blakie2023,Ilg2023gss,Roccuzzo2019,Sindik2023sound,Platt2024}. In an axially infinite geometry, each branch can be assigned a character: one branch corresponds to phase excitations, i.e.~excitations inducing density exchange between crystal sites without motion of the sites themselves; the other corresponds to crystal (also called density) excitations, i.e.~excitations inducing changes in the crystal structure without altering the sites' population. Both branches have a linear dispersion $\epsilon(q) \sim_{q\rightarrow 0} c_\nu q$ where $c_\nu$ is the speed of sound associated with the branch $\nu=1,2$ (phase and crystal sounds, also called second and first sounds). For dipolar gases, the phase branch has a lower energy than the crystal branch and therefore a lower sound speed. Furthermore, the crystal sound speed increases with decreasing $\as$ while the phase sound speed decreases, evidencing the reduction in superfluid fraction, see also \ref{par:furtherremarks}.

A regime of particular interest in tubular supersolids is that of intermediate densities, where the superfluid-supersolid transition is continuous, see \ref{par:tubesupersolids}. We have seen that the supersolid here connects to the softened roton mode. This connection extends beyond the state itself, and the full excitation spectrum of the supersolid links to the rotonic spectrum of the unstable superfluid. The mapping is done by folding the superfluid spectrum onto an effective Brillouin zone defined by the periodicity of the roton mode, i.e. with a width $\krot/2$, see Fig.\,\ref{fig:spectra} (II.a)~\cite{Blakie2023,Hertkorn2019fot}. Through this folding, the crystal excitation branch connects to the gapless branch at $k\rightarrow 0$, while the phase branch links to that at $k\rightarrow \krot^-$. Meanwhile, the branch that connects to $k\rightarrow \krot^+$, which is also gapless at $\as=\ac$, experiences a gap opening for decreasing $\as<\ac$, see Fig.\,\ref{fig:spectra} (II.b). This branch corresponds to the Higgs mode~\cite{Pekker2015ami,Hertkorn2019fot}. As its gap starts to open, the Higgs branch mixes with the gapless branches in the vicinity of the phase transition. This mixing results in both a strong damping of the Higgs mode~\cite{Hertkorn2019fot} and a discontinuity of the speeds of sound of the Goldstone branches at the transition~\cite{Blakie2023}. The discontinuity of the speeds of sound also underlies a discontinuity of the system's compressibility and indicates the transition's second-order character. At the density where $\ac$ is maximal, the discontinuity vanishes, indicating a higher-order transition at this point.

Similar features of the supersolid excitation spectrum survive in finite harmonically trapped systems despite the discrete nature of their spectrum and the fact that quasimomentum is no longer a good quantum number. Two phononic branches could be identified theoretically and experimentally, yet the modes' characters mix~\cite{Natale2019eso,tanzi2019ssb,guo2019tle}. An amplitude mode could also be pinpointed in Bogoliubov calculations. Remarkably, this mode is stable in the direct vicinity of the transition due to the discreteness of the spectrum. Yet, further away, it also hybridizes with other modes analogously to infinite systems~\cite{Hertkorn2019fot}. A recent theoretical work~\cite{Hertkorn2024dsa} suggests that in finite toroidal traps, Higgs and Goldstone modes decouple due to the absence of coupling between modes of different quasimomenta, which corresponds to distinct circulation numbers in the finite system.

\subsubsection{Pancake geometries}
For supersolids with bidimensional crystals ($D=2$), such as those formed in pancake geometries and reviewed in \ref{par:pancakesupersolids}, the excitation spectrum is both richer and changes with the crystal structure. In infinite pancakes, three Goldstone modes are present: Two of them are longitudinal modes where motion occurs along the excitation propagation direction. These two modes are similar to those of $D=1$ supersolids, and are associated with crystal and phase excitations, respectively. The remaining Goldstone mode is a shear mode where crystal motion occurs transversely to the wave propagation direction~\cite{Poli2024eoa,Blakie2025dpa}. All three Goldstone branches exhibit linear dispersion for $q \rightarrow 0$ with distinct speeds of sound $c_\nu$, $\nu=1,2,3$.

For all crystal configurations -- triangular, stripe, honeycomb -- of $D=2$ supersolids, the longitudinal crystal branch (first-sound branch), has higher energy and sound speed than the two other branches~\cite{Poli2024eoa,Blakie2025dpa}. This first sound speed weakly increases with decreasing $\as$. Just below the uniform-to-supersolid transition, i.e. at $\as\lesssim \ac$, the shear mode has lower energy and sound speed than the longitudinal phase mode (second sound).  

The further behavior of shear and phase branches depends on the crystal configuration. For triangular supersolids, the shear sound speed increases with decreasing $\as$ while the second sound speed decreases, similarly to $D=1$ supersolids~\cite{Poli2024eoa}. At low $\as$,  the second sound becomes the lowest branch until it fully softens, indicating the transition to an insulating crystal, see Fig.\,\ref{fig:spectra} (III.a). Conversely, for honeycomb supersolids, both shear sound and second sound speeds decrease with decreasing $\as$, so that the shear branch remains the lowest one~\cite{Blakie2025dpa,Yapa2024ssw}, see Fig.\,\ref{fig:spectra} (III.b). The softening of this shear mode with decreasing $\as$ underlie a transition to the stripe state where shear excitations have no energy cost. 
As discussed in \ref{par:pancakesupersolids}, the honeycomb to stripe is of first order and the shear instability occurs at a lower $\as$ than the transition itself, underlying the system's bistability. 

The spectrum of $D=2$ supersolids in finite pancake traps has also been explored, focusing on the triangular case~\cite{Hertkorn2021sit}.  Here a trap of moderate aspect ratio $\omega_\perp/\omega_\parallel=2$ is considered for which the MF instability is driven by an angular roton, see \ref{par:fulltrapexc}. The softening of the angular roton gives rise to a zero-energy Goldstone mode corresponding to the breaking of the rotation symmetry and to a Higgs mode. As for $D=1$ supersolids, the Higgs mode hybridizes with other modes as $\as$ is lowered and the Higgs gap opens. 

\subsubsection{Further remarks}\label{par:furtherremarks}
For both $D=1$ and $D=2$ supersolids, hydrodynamic theories have been developed, describing the system's behavior based on broken symmetries and conservation relations~\cite{Son2005,Josserand2007pas, Yoo2010, Hofmann2021,Sindik2023sound,Poli2024eoa, Platt2024, Platt2024ss}. These theories permit to relate the different speeds of sound to the superfluid fraction, compressibility, and elasticity. Through these relations, a non-vanishing second sound (i.e.~phase sound) speed implies a finite superfluid fraction and thus the superfluid character of the supersolid. Furthermore, key parameters such as compressibility and superfluid fraction can be extracted by measuring all sound speeds, offering a promising experimental approach to gain insights onto the supersolid's superfluid character and dynamical behavior~\cite{Biagioni2024mot,Platt2024ss,Sindik2023sound}. 

Beyond their elementary excitations, other intriguing excitations of superfluids -- and supersolids -- are those arising as a response to rotational perturbations. Being irrotational, these systems are expected to accommodate vortices, i.e.~localized defects with a phase winding of $2\pi \times p$, $p \in \mathbb{Z}$. This phase winding implies that each particle carries $p$ quanta of angular momentum.   Vortices have been both predicted and observed in $D=2$ supersolids under rotational perturbations~\cite{gallemi2020qvi,Rocuzzo2020ras,Ancilotto2021vpi,Casotti2024obov}. This provides additional proof of their superfluid behavior and opens new avenues for exploring their dynamics.

\section{Conclusions and Perspectives}\label{sec:conclusion}

In this review, we have examined the rich variety of ground states stabilized beyond the MF instability in dipolar quantum gases made of magnetic atoms.
Since the discovery of the stabilization mechanism based on quantum fluctuations a decade ago, extensive experimental and theoretical efforts have unveiled the remarkable properties of these states and deepened our understanding. 
Breakthroughs have been achieved, including the first observation of spontaneously crystallized states~\cite{Kadau2016otr}, the formation of ultradilute macrodroplets~\cite{Chomaz2016qfd,schmitt2016sbd}, the observation of both supersolid and insulating modulated states in tubular and pancake geometries~\cite{Chomaz2019lla,Tanzi2019ooa,Bottcher2019tsp,Norcia2021tds}, the probing of the remarkable feature of their spectrum of elementary excitations~\cite{Natale2019eso,guo2019tle,tanzi2019ssb} as well as their answer to various perturbations, including rotation~\cite{Tanzi2021eos,Casotti2024obov}.

Despite these achievements, many exciting directions remain open for exploration.  Further exploration of the phase diagram may reveal supersolids with exotic structures, including stripe, honeycomb, labyrinth, and pumpkin configurations~\cite{Hertkorn2021pfi, Ripley2023Two}. Beyond the characterization of the individual phases, the transitions between these phases and their associated critical properties provide a rich landscape for advancing our understanding of complex quantum many-body behavior~\cite{Kirkby2024kzs, Blakie2025dpa,Yapa2024ssw}.  Moreover, the investigation of the dynamical and hydrodynamical properties of these exotic states is still in its infancy. In particular, much remains to be discovered about the interplay between crystalline order and superfluidity in supersolids~\cite{gallemi2020qvi,Rocuzzo2020ras,Ancilotto2021vpi,Casotti2024obov,Biagioni2024mot,Platt2024ss,Sindik2023sound,Blakie2025dpa,Yapa2024ssw,Hertkorn2021sit,Poli2024eoa,Hertkorn2019fot,Tanzi2021eos,Ilzhofer2021pci}.  The realization of exotic traps, and in particular toroidal traps, offers a promising path to investigate these behaviors in purer environments, allowing the decoupling of the different supersolid excitations and the establishment of persistent flow~\cite{Hertkorn2024dsa}.
 Understanding the supersolid behaviors may provide insights that extend well beyond the dipolar gases themselves. For instance, analogies of supersolids with certain regions of neutron stars have been suggested, opening up intriguing prospects for quantum simulations of these cosmological systems~\cite{Poli2023gir}.

Alongside the rapid progress with magnetic atoms, experimental and theoretical studies have expanded to other dipolar systems. Notably, quantum gases of polar molecules and mixtures with at least one dipolar species have seen significant experimental advancements in recent years~\cite{Bigagli2024oob, Trautmann2018dqm, Lecomte2025pas}. These developments have sparked predictions of novel behaviors in quantum-stabilized states within these alternative platforms~\cite{Schmidt2022sbd,Kirkby2024eoa}.

The impact of the magnetic-atom results extends even beyond dipolar gases. They reignited interest in supersolid-like phases in a variety of platforms, including spin-orbit coupled BECs~\cite{Li2017asp, putra2020spatial,chisholm2024pst}, driven BECs~\cite{liebster2025oos}, exciton-polariton~\cite{trypogeorgos2025esi}, and certain condensed-matter systems~\cite{xiang2024gme,Conti2023cso}. This expanding landscape promises rich cross-fertilization between platforms and paves the way for future discoveries.

\begin{acknowledgments}

We thank Danny Baillie, Blair Blakie, Karthik Chandrashekara, Christian Gölzhaüser, Lennart Hoenen, Felix Kaufmes, Wyatt Kirkby, and Philipp Lunt for their careful reading and insightful comments on this manuscript. This work is funded by the European Research Council (ERC) under the European Union’s Horizon Europe research and innovation program under grant number 101040688 (project 2DDip), and by the Deutsche Forschungsgemeinschaft (DFG, German Research Foundation) through project-ID
273811115 (SFB1225 ISOQUANT) and under Germany’s Excellence Strategy EXC2181/1-390900948 (the Heidelberg Excellence Cluster STRUCTURES). 
\end{acknowledgments}
\vspace{5pt}

\appendix
\renewcommand\thefigure{\thesection S\arabic{figure}}   
\setcounter{figure}{0}

\end{document}